\documentclass[sn-mathphys,Numbered,fleqn,10pt]{wlscirep}

\usepackage[utf8]{inputenc}
\usepackage[T1]{fontenc}
\usepackage{wasysym}
\usepackage{braket,bm} 
\usepackage{amssymb}
\usepackage{dsfont}
\usepackage{bbold}
\usepackage{graphicx}
\usepackage{float}
\usepackage{ulem}
\usepackage{comment}
\usepackage{soul}

\title{Qubit Analog with Polariton Superfluid in an Annular Trap}

\author[1,2,3]{J. Barrat}
\author[4,5]{A. F. Tzortzakakis}
\author[1,2,3]{M. Niu}
\author[2,3]{X. Zhou}
\author[2,3]{G.G. Paschos}
\author[4]{D. Petrosyan}
\author[2,3,4]{P.G. Savvidis}

\affil[1]{Department of Physics, Fudan University, Shanghai 200433, China}
\affil[2]{Key Laboratory for Quantum Materials of Zhejiang Province, Physics Department, Westlake University, 18 Shilongshan Rd, Hangzhou 310024, Zhejiang, China}
\affil[3]{Institute of Natural Sciences, WIAS, 18 Shilongshan Road, Hangzhou, Zhejiang Province 310024, China}
\affil[4]{Institute of Electronic Structure and Laser, FORTH, 70013 Heraklion, Crete, Greece}
\affil[5]{Department of Physics, National and Kapodistrian University of Athens, 15784 Athens, Greece}

\twocolumn

\begin{abstract}

We report on the experimental realization and characterization of a qubit analog with semiconductor exciton-polaritons. In our system, a condensate of exciton-polaritons is confined by a spatially-patterned pump laser in an annular trap that supports energy-degenerate circulating currents of the polariton superfluid. Using temporal interference measurements, we observe coherent oscillations between a pair of counter-circulating superfluid vortex states of the polaritons coupled by elastic scattering off the laser-imprinted potential.
The qubit basis states correspond to the symmetric and antisymmetric superpositions of the two vortex states forming orthogonal double-lobe spatial wavefunctions. By engineering the potential, we tune the coupling and coherent oscillations between the two circulating current states, control the energies of the qubit basis states, and initialize the qubit in the desired state. The dynamics of the system is accurately reproduced by our theoretical two-state model, and we discuss potential avenues to achieve complete control over our polaritonic qubits and realize controllable interactions between such qubits to implement quantum gates and algorithms analogous to quantum computation with standard qubits.

\end{abstract}

\begin{document}

\maketitle


\paragraph{Introduction.}

Exciton-polaritons are hybrid light-matter quasiparticles resulting from the strong coupling of semiconductor excitons and microcavity photons\cite{Deng_Exciton_2010,carusotto2013quantum}. The polaritons combine an extremely small effective mass, inherited from their photonic component, with strong nonlinearities, inherited from their excitonic component. As a result, polaritons can exhibit macroscopic spatial coherence and form out-of-equilibrium condensates exhibiting superfluid behavior at elevated temperatures when pumped above threshold \cite{Deng_Condensation_2002,Kasprzak_Bose_2006}. Furthermore, the dissipative nature of polaritonic condensates, primarily stemming from their photonic component, enables interferometric measurements of condensate luminescence and extraction of the macroscopic wavefunction while still maintaining long coherence times.  
Given their unique properties, exciton-polaritons represent an attractive platform for exploring quantum collective phenomena and hold promise for a variety of applications, including low-threshold lasing\cite{Imamoglu_Nonequil_1996,Christopoulos_Room_2007}, efficient energy transfer \cite{Brien_Photonic_2009,Sanvitto_Theroad_2016}, and simulation of many-body systems \cite{Amo_Exciton_2016,Kim_Exciton_2017}.

A promising recent theoretical proposal for quantum computing utilizes split-ring polariton-condensate qubits involving quantized circular currents\cite{Kavokin_split_ring_qubit, Kavokin_polar_q_comput}. This system relies on the formation of vortices in superfluids arising from the quantization of circulation, where the phase accumulation around a supercurrent loop can only take discrete values. Closely related physics governs the principles of operation of superconducting flux or phase qubits involving superconducting loops interrupted by Josephson junctions\cite{Mooij_Josephson_1999,Chiorescu_Coherent_2003,You_Atomic_2011,Arute_Quantum_2019}. 

Following their initial observation in planar unconfined geometries \cite{Lagoudakis_1,Lagoudakis_2,Manni_Dissociation_2012,Grosso_Soliton_2011,Nardin_Hydrodynamic_2011},  quantized vortices and persistent circulating currents of exciton-polariton superfluids have been observed and studied in various ring-shaped geometries, including optically-induced traps \cite{Dall_Creation_2014,Gao_Chiral_2018,Gao_Controlled_2018,Sedov_Circular_2021,Aladinskaia_Spatial_2023} or confinement in etched micropillars \cite{Real_Chiral_2021,Lukoshin_Persistent_2018}. 
Previous efforts to generate polariton condensates with nonzero orbital angular momentum (OAM) employed phase-shaped external laser beams \cite{Sanvitto_Persistent_2010,Kwon_Direct_2019,Dominici_Shaping_2021,Choi_Observation_2022} or optically generated nearly defect-free potentials \cite{Ma_Realization_2020,Alyatkin_All}. These techniques have successfully overcome the limitations imposed by intrinsic potential disorder which often leads to phase-locking and the formation of standing waves manifested by petal-shaped patterns pinned by the defects \cite{Manni_Spontaneous_2011,Dreismann_Coupled_2014,Wang_Spontaneously_2021,Zhang_All_2023}. Moreover, slightly detuned Laguerre-Gaussian (LG) beams with nonzero OAM have been utilized to stir the trapping potential and control the vorticity of the superfluid \cite{Gnusov_Quantum_2023}. So far, however, none of these approaches have achieved controllable superpositions of vortex states. 

Here we show that, under appropriate conditions, optically trapped non-equilibrium polariton condensates can populate two well-characterized states corresponding to the clockwise and counterclockwise circulating currents. We demonstrate coherent coupling between these states, due to the partial reflection of the circulating superfluid from a weakly disordered laser potential or an external control laser beam, while simultaneously maintaining long coherence times. 
We can control the coupling and thereby the energy splitting between the two eigenmodes of the system. Inspired by the theoretical proposal to realize qubit analogs and quantum computing with two-mode BECs\cite{Byrnes2012}, we formally identify the two polaritonic eigenmodes with the basis states of a qubit.  
Supplemented with controllable coupling between individual polaritonic qubits, such systems hold great potential for simulating a subset of quantum algorithms that do not rely on entanglement \cite{Biham2004,Lanyon2008,Balynsky2021,Mohseni2022}. 

\begin{figure}[t]
    \includegraphics[clip,width=1.0\linewidth]{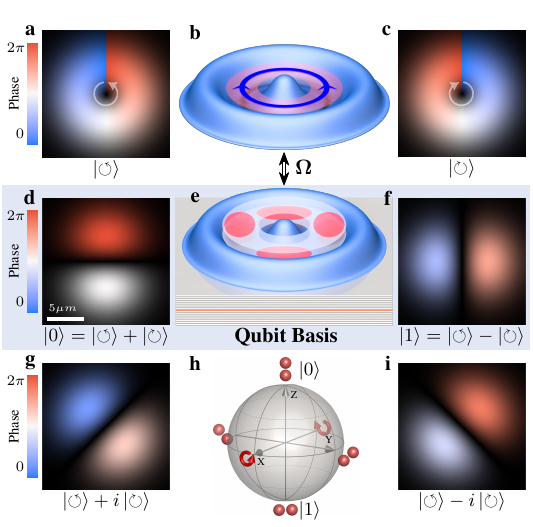}
    \caption{\textbf{Schematics of the polaritonic qubit analog.} 
    A pump laser creates a Mexican hat-shaped potential for the polaritonic condensate (\textbf{b}).
    The potential supports two energy-degenerate counter-circulating polariton modes $\ket{\circlearrowleft}, \ket{\circlearrowright}$ (\textbf{a}, \textbf{c}), which are coupled with rate $\Omega$ by small defects or ellipticity of the trapping potential. The resulting superpositions $(\ket{\circlearrowleft}\pm \ket{\circlearrowright})/\sqrt{2}$ of the two vortex modes form vertically and horizontally oriented two-lobe eigenmodes (\textbf{d},\textbf{f}) that represent the qubit basis states $\ket{0,1}$ energy-split by $\mp\Omega$ (\textbf{e}). 
    Another pair of states $e^{\pm i\pi/4}(\ket{\circlearrowleft}\mp i\ket{\circlearrowright})/\sqrt{2}$ with diagonally oriented lobes (\textbf{g},\textbf{i}) represents the third axis of the Bloch sphere (\textbf{h}) for the continuum of states of the system. In panels (\textbf{a}, \textbf{c}, \textbf{d}, \textbf{f}, \textbf{g}, \textbf{i}) the shading is proportional to the amplitude of the polaritonic condensate and the color encodes its phase varying from $0$ to $2\pi$.
    }
    \label{fig:figure1}
    \end{figure}

\paragraph{Polaritonic qubit analog}.
Our system is illustrated in Fig.~\ref{fig:figure1}. A spatially-patterned pump laser creates a Mexican hat-shaped trapping potential of appropriate size for the polaritons Fig.~\ref{fig:figure1}(b). When pumped slightly above the condensation threshold, the polaritons condense into two energy-degenerate counter-circulating vortex states $\ket{\circlearrowleft}$ and $\ket{\circlearrowright}$ whose wavefunctions can be well approximated by the Laguerre-Gaussian modes with orbital angular momenta $l=\pm 1$: 
\begin{equation}
    \psi_{\circlearrowleft,\circlearrowright}(\bm{r})= C\, \rho \,e^{-\frac{1}{2} (\rho/\rho_c)^2}\, e^{\pm i \theta}, \label{eq:Psivortex}
\end{equation}
where we use cylindrical coordinates $\bm{r} \equiv (\rho,\theta,h=0)$, $\rho_c$ is the mean radius of the condensate, and $C$ is a normalization constant.

Small defects and weak ellipticity of the confining potential result in backscattering of the polariton current and thereby coherent coupling between the states $\ket{\circlearrowleft}$ and $\ket{\circlearrowright}$ with rate $\Omega$. 
We may formally associate our system with a spin-1/2 system via mapping 
$\ket{\circlearrowleft,\circlearrowright}\to \ket{\uparrow,\downarrow}_x$,  
$e^{\pm i\pi/4} \dfrac{\ket{\circlearrowleft}\mp i \ket{\circlearrowright}}{\sqrt{2}}\to\ket{\uparrow,\downarrow}_y$,
and $\dfrac{\ket{\circlearrowleft}\pm\ket{\circlearrowright}}{\sqrt{2}}\to\ket{\uparrow,\downarrow}_z$
and write the Hamiltonian for the two coupled vortex states as ($\hbar =1$)
\begin{equation}
    H= -\Omega \ket{\circlearrowleft} \bra{\circlearrowright} + \mathrm{H.c}  = -\Omega \sigma_z , \label{eq:Ham}
\end{equation}
where $\sigma_z = \ket{\uparrow}_z \bra{\uparrow} - \ket{\downarrow}_z \bra{\downarrow}$ is the Pauli spin operator while $\Omega$ plays the role of the effective longitudinal magnetic field (along the $z$ direction) that results in energy splitting $\mp \Omega$ between the states $\ket{\uparrow}_z$ and $\ket{\downarrow}_z$ shown in Fig.~\ref{fig:figure1}(e). We can thus assign to the lower and higher energy states the qubit basis states $\ket{0,1} \equiv \ket{\uparrow,\downarrow}_z$.
The continuum of states of the system is conventionally represented by the Bloch sphere, see Fig.~\ref{fig:figure1}(h).

The spatial wavefunctions of the states $\ket{\uparrow,\downarrow}_{x,y,z}$ are shown in Fig.~\ref{fig:figure1}(a,c), (g,i), (d,f), respectively. We observe that the vortex states $\ket{\uparrow,\downarrow}_x =\ket{\circlearrowleft,\circlearrowright}$ have azimuthally uniform amplitude and phase increasing from $0$ to $2\pi$ in counterclockwise and clockwise directions, whereas the vortex superposition states $\ket{\uparrow,\downarrow}_y$ and $\ket{\uparrow,\downarrow}_z$ have two lobes with diagonal and vertical/horizontal orientations with the phase difference $\pi$ between the lobes. With the convention that the potential energy bump or minor axis of the ellipse of the confining potential is oriented horizontally, the eigenmode $\ket{\uparrow}_z$ with the vertically oriented lobes has lower energy and the eigenmode $\ket{\downarrow}_z$ with horizontally oriented lobes has higher energy.


\paragraph{Experimental setup.}
The experiments are carried out at a cryogenic temperature of 11K in a semiconductor microcavity consisting of 4 sets of 3 GaAs quantum wells placed at an antinode of a high finesse cavity formed by AlGaAs/AlAs Bragg mirrors. The sample is excited by a continuous wave laser, pumping a reservoir of excitons, some of which relax to form polaritons. The excitonic reservoir also induces a repulsive potential for the polaritons via interaction with their excitonic component. Details of the experimental setup are given in Supp. Mat.~\ref{sec:app:experiment}.

\begin{figure}[t]
    \includegraphics[clip,width=1\linewidth]{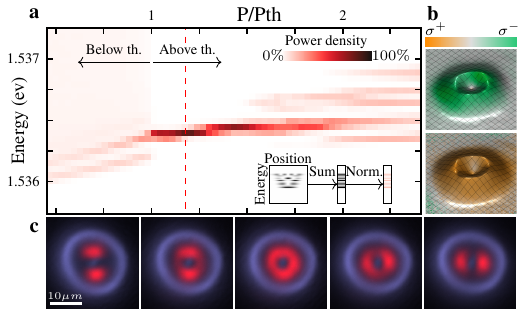}
    \caption{\textbf{Polariton condensate in an annular trap.}
    \textbf{a}.~Integrated photoluminescence spectrum of polaritons vs the pumping strength $P$, with inset illustrating the schematics of the measurement. 
    Slightly above the condensation threshold $P=1.2P_\mathrm{th}$ (dashed vertical line), the polaritons condense to the vortex states with orbital angular momenta $l=\pm 1$ having the same energy. 
    \textbf{b}.~Polariton condensate intensity (profile height) and spatially-resolved degree of circular polarization (color) for $\sigma^-$ (top) or $\sigma^+$ (bottom) polarized pump laser. The circularly polarized pump generates polariton condensate with the same spatially-uniform polarization. 
    \textbf{c}.~Polariton condensate photoluminescence (red) for increasing (from left to right) ratios of the horizontal and vertical axes of the elliptic profile of the pump laser (blue).}
    \label{fig:figure2}
\end{figure}

The pump laser is spatially patterned using a spatial light modulator (SLM) to imprint a Mexican hat-shaped potential with a radius of $\sim 9\;\mu$m to confine the polaritons. In such a small trap, just above the condensation threshold $P\approx 1.2\, P_{\text{th}}$ for the polaritons, we observe only a narrow energy mode, while for stronger pumping $P > 1.5\, P_{\text{th}}$ the reservoir of excitons can populate many discreet polariton modes. This is verified by measuring the angle-integrated photoluminescence intensity and extracting the normalized power density of spectrometer CCD (energy resolution per pixel is $45\;\mu$eV), as shown in Fig.~\ref{fig:figure2}(a).

In what follows, we consider the single-mode regime of weak pumping $P\approx 1.2\, P_{\text{th}}$. Inside the trap, the bosonic polaritons form an annular condensate with a mean radius $\rho_c \simeq 4\,\mu$m, spatially separated from the reservoir excitons injected by the pump, see Fig.~\ref{fig:figure2}(c).  
The trapping also suppresses the spin-obit coupling of the polaritons \cite{Mukherjee_Dynamics_2021,Liu_Anew_2015,Dominici_Vortex_2015,Gnusov_All_2021,Pukrop_Circular_2020,Demirchyan_Spinor_2022,Aristov_Screening_2022}, and we observe a  high degree of polarization of the condensate\cite{Gnusov_Optical_2020} strongly correlated with the polarization $\sigma^+$ or $\sigma^-$ of the pumping laser, see Fig.~\ref{fig:figure2}(b).

\begin{figure}[t!]
    \includegraphics[width=1\linewidth]{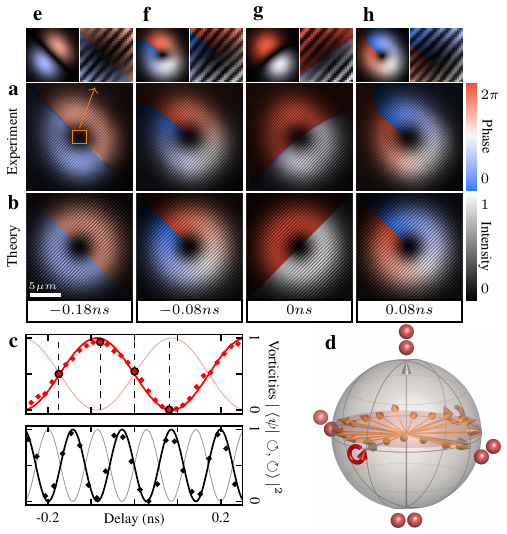}
    \caption{\textbf{Interferometric measurement of the system dynamics.} Time-averaged spatial interferograms of the polaritonic condensate at four different delay times $\tau = -0.18,-0.08,0,0.08\;$ns between the signal beam and the reference taken at $\rho_0 \simeq \rho_c$ and $\theta_0 \simeq \pi/4$, as obtained from the experimental measurements (\textbf{a}) and theoretical model (\textbf{b}).
    The condensate amplitude extracted from the interferometric images and the magnified view of the center of interferograms reveal the presence or absence of singularity (fork-shaped interference fringes) associated with the vortex states (panels \textbf{e}, \textbf{f}, \textbf{g}, \textbf{h}). 
    The polariton wavefunction $\psi$ exhibits continuous oscillations with frequency $\Omega$ between the two vortex states $\ket{\circlearrowleft}$ and $\ket{\circlearrowright}$, quantified by vorticities $|\braket{\psi | \circlearrowleft,\circlearrowright}|^2$.
    Introducing a potential barrier that enhances the scattering between the vortex states increases the oscillation frequency (lower panel in \textbf{c}). 
    On the Bloch sphere, the dynamics of the system corresponds to the spin precession in the $xy$ plane slightly biased towards $z$ direction (ground state $\ket{0}$) (\textbf{d}).}
    \label{fig:figure3}
\end{figure}

A nearly perfectly circular trapping potential imposed by the pump laser supports a ring-shaped polariton condensate consisting of two coupled counter-rotating superfluid vortex modes, as described above. Although presently there is no consensus on unambiguous criteria for superfluidity in non-equilibrium systems\cite{juggins_2018}, we note that the persitent circulation of polariton fluid under non-resonant excitation and the emergence of quantized vortices provide compelling evidence of superfluid behavior\cite{Kwon_Direct_2019,Choi_Observation_2022}. When the pump laser profile is elliptic, we observe the double-lobe (standing-wave) structure oriented along the major axis of the ellipse, see Fig.~\ref{fig:figure2}(c). We therefore engineer the spatial profile of the pump laser via precise control of the SLM to minimize the ellipticity of the trapping potential.


\paragraph{Dynamics of the system.}
The polariton condensate governed by Hamiltonian (\ref{eq:Ham}) undergoes Rabi-like oscillations between the two vortex states $\ket{\circlearrowleft}\equiv\ket{\uparrow}_x$ and $\ket{\circlearrowright}\equiv \ket{\downarrow}_x$ with frequency $\Omega$ (see Figs.~\ref{fig:figure3}) and the state of the system at any time $t$ can be cast as 
\begin{equation}
    \ket{\psi(t)} = \cos(\Omega t +\phi_0)\ket{\circlearrowleft} + i \sin (\Omega t +\phi_0)\ket{\circlearrowright},
\end{equation}
where $\phi_0$ is some initial phase of the oscillations. This dynamics can be visualized as a precession of the spin in the $xy$ plane about the effective magnetic field $\Omega$ along the $z$ direction. 

To resolve experimentally the dynamics of the system, we perform interferometric measurements of the polariton photoluminescence. The light emanating from the polaritonic condensate $\psi(\bm{r},t)$  is split by a beamsplitter and sent into two arms of an interferometer. 
In one arm, the signal beam passes through a variable delay line to introduce time delay $\tau$. In the other arm, the beam is expanded and the field at position $\bm{r}_0 = (\rho_0,\theta_0)$ is taken as a reference with nearly uniform amplitude and phase front in the plane perpendicular to the propagation direction. The two beams are combined on a CCD camera with long exposure time $T$, resulting in interferometric image   
\begin{equation}
    \langle\, I(\bm{r}, \tau)\, \rangle \propto \dfrac{1}{T} \int_0^{T} \big|\psi(\bm{r},t + \tau) e^{i\bm{k}_{\mathrm{S}} \cdot \bm{r}} + \psi(\bm{r}_0,t) e^{i\bm{k}_{\mathrm{R}} \cdot \bm{r}} \big|^2 ,  
\end{equation}
where $\bm{k}_{\mathrm{S,R}}$ are the wavevectors of the signal and reference beams and $\Delta \bm{k}\equiv \bm{k}_{\mathrm{R}}-\bm{k}_{\mathrm{S}}$ determines the spatial separation of the interference fringes.

In Fig.~\ref{fig:figure3}(a,b) we show the experimentally obtained interferograms for $\rho_0 \simeq \rho_c$ and $\theta_0 \simeq \pi/4$ at different time delays $\tau$, and compare them with the theoretical model revealing excellent agreement. By performing Fourier analysis \cite{Real_Chiral_2021,Sigurdsson_Persistent_2022} of the images $\langle\, I(\bm{r}, \tau)\, \rangle$ (see Supp. Mat.~\ref{sec:app:interference} for details), we can retrieve the polariton wavefunction as 
\begin{equation}
    \mathcal{I}_{-\Delta\bm{k}}(\bm{r},\tau)=e^{-i\pi/4}\,\psi(\bm{r},t=\tau) \quad (\phi_0=\pi/4), 
\end{equation}
where $\mathcal{I}_{-\Delta\bm{k}}$ is the corresponding Fourier component centered around $-\Delta\bm{k}$. The wavefunction oscillates between the $\psi_{\circlearrowleft}(\bm{r})$ and $\psi_{\circlearrowright}(\bm{r})$ vortex modes with period $\Delta \tau = \pi/\Omega$, see Fig.~\ref{fig:figure3}(c). 
Using an additional weak control beam to introduce a potential barrier that reflects the polariton currents, we can thereby enhance the coupling between the two vortex states leading to oscillations with increased frequency $\Omega$.

\begin{figure}[t!]
    \includegraphics[clip,width=1\linewidth]{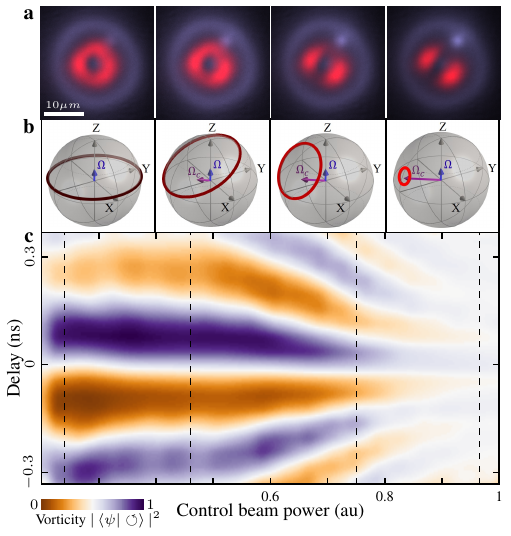}
    \caption{\textbf{Oscillation control and state initialization.} \,
    \textbf{a}. Polariton condensate photoluminescence (red) in a laser trap (light blue) for increasing intensity of the control beam (bright blue spot). 
    A weak control beam contributes to the coupling between the two vortex states $\ket{\circlearrowleft}$ and $\ket{\circlearrowright}$, while a strong control beam pins the condensate to a standing-wave with a node at the beam position, i.e., the two lobes avoiding the beam. 
    \textbf{b}.~Oscillation trajectories (one period) of the effective spin on the Bloch sphere with the original orientation of the axes for four different intensities of the control beam as in \textbf{a}. 
    The position and strength of the control beam define the direction and magnitude of the effective magnetic field $\Omega_c$ that adds to the intrinsic field $\Omega$. 
    \textbf{c}.~Condensate oscillation dynamics vs the control beam power. For a very weak beam, the oscillation period $\pi/\Omega \simeq 120\;$ps between the two vortex states is determined by the effective magnetic field $\Omega$ due to the intrinsic disorder of the trapping potential. Stronger control beam leads to an increase of oscillation frequency $\Omega_{\mathrm{tot}}$ and suppression of population of the higher energy eigenstate $\ket{\downarrow}_\zeta$ with one of the lobes on top of the potential energy bump induced by the beam, which decreased the oscillation amplitude.  
    }
    \label{fig:figure4}
\end{figure}

In Fig.~\ref{fig:figure4} we illustrate in greater detail the influence of a tightly focused control beam on the dynamics of the system. A weak control beam creating a potential bump smaller than the intrinsic defects of the trap does not change noticeably the oscillation frequency between the two vortex states $\ket{\circlearrowleft}$ and $\ket{\circlearrowright}$. But increasing the control beam intensity and thereby the height of the potential barrier enhances the scattering of the polaritons between the counter-circulating vortex modes which in turn leads to faster oscillations with decreasing amplitude. For very strong control beam, the condensate is pinned to a standing-wave (double-lobe) mode with a node at the control beam position. 

In the spin language, the control beam at any position of the annular trap corresponds to an effective magnetic field $\Omega_c$ pointing in some direction $\hat{d}$ in the $yz$ plane. The total effective magnetic field $\hat{\zeta} \Omega_\mathrm{tot} = \hat{z} \Omega + \hat{d} \Omega_c$ is given by the vector sum of the intrinsic $\hat{z} \Omega$ and applied $\hat{d} \Omega_c$ fields. 
The spin precesses about the total magnetic field $\hat{\zeta} \Omega_\mathrm{tot}$ which thus defines the quantization direction $\hat{\zeta}$ and the two eigenstates $\ket{\uparrow}_\zeta$ and $\ket{\downarrow}_\zeta$ corresponding to the standing-wave (double-lobe) modes with orthogonal orientation. These two eigenstates are split by $\mp \Omega_{\mathrm{tot}}$ and the dynamics of the system is given by
\begin{equation}
\ket{\psi (t)} = c_0 e^{i \Omega_{\mathrm{tot}} t} \ket{\uparrow}_\zeta + c_1 e^{-i \Omega_{\mathrm{tot}} t} \ket{\downarrow}_\zeta,
\end{equation}
where $|c_{0,1}|^2$ are the stationary populations of the two eigenstates. For equal populations $|c_0|^2 = |c_1|^2$, the spin processes in the plane perpendicular to $\hat{\zeta}$ passing through the two vortex states $\ket{\circlearrowleft} = (\ket{\uparrow}_\zeta + \ket{\downarrow}_\zeta)/\sqrt{2}$ and $\ket{\circlearrowright}= (\ket{\uparrow}_\zeta - \ket{\downarrow}_\zeta)/\sqrt{2}$. 
But with increasing imbalance, $|c_0|^2 > |c_1|^2$, these oscillation amplitude is decreasing, and for $|c_0|^2 \gg |c_1|^2$ the spin is pinned to the lower energy state $\ket{\uparrow}_\zeta$.

In the polariton picture, the stronger control beam leading to reflection of the polariton current increases the energy separation between the two standing-wave eigenmodes. But the same localized potential barrier simultaneously increases the polariton scattering in all directions. This scattering, or extra loss, affects only the higher energy eigenmode with the antinode at the potential barrier, which induces imbalance of populations $|c_0|^2 \gg |c_1|^2$ (see Supp. Mat. \ref{sec:app:popdiff}).

\begin{figure}[t]
    \includegraphics[clip,width=1\linewidth]{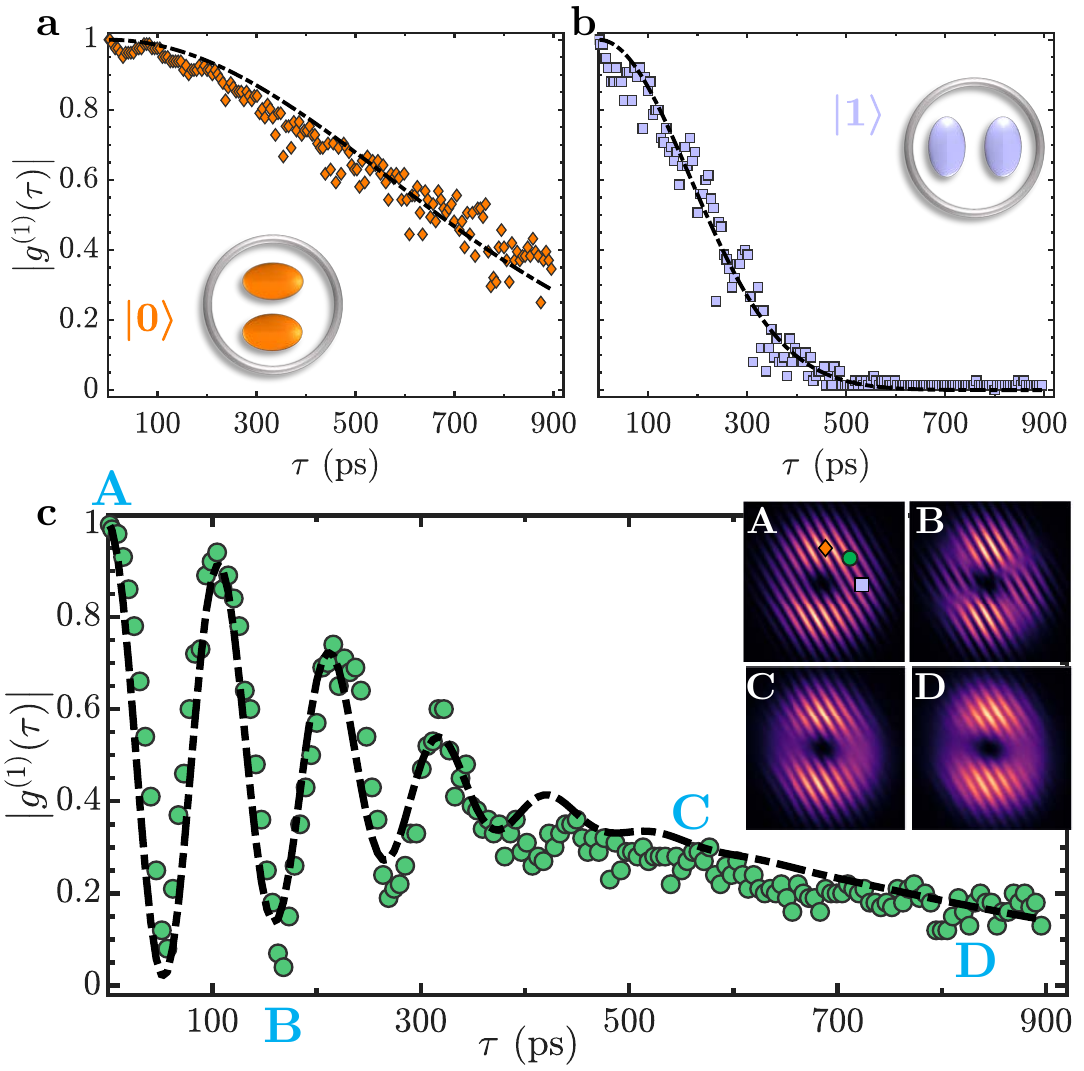}
    \caption{\textbf{Coherence measurements.} Time-dependence of the absolute value of the first-order correlation function $|g^{(1)}(\tau)|$ as obtained from the experimental measurements (filled circles) and fitted with the theoretical two-mode model (dashed lines) yielding the coherence times $\tau_{c0} \simeq 260\,$ps and $\tau_{c1} \simeq 800\,$ps. 
     Correlation functions of the lower (\textbf{a}) and higher (\textbf{b}) energy states $\ket{0}$ and $\ket{1}$, and  
     correlation function for the superposition of the two states (\textbf{c}). 
    The insets in (\textbf{c}) show the corresponding interferograms at time delays $\tau \simeq 0, 170, 580$, and 805 ps, denoted as A, B, C, and D. 
    In the upper left inset, filled markers of different colors show the positions of the interferometric measurements of $|g^{(1)}(\tau)|$ for graphs (\textbf{a}, \textbf{b}, \textbf{c}).}
    \label{fig:figure5}
\end{figure}

\paragraph{Coherence measurements.}
The above experimental and theoretical analysis is relevant for times shorter than the coherence time $\tau_c$ of the system. We now employ a Mach-Zehnder interferometer to perform measurements of the first-order correlation function of the polariton condensate 
\begin{equation}
    g^{(1)}(\bm{r},\tau) \propto \lim_{T\to\infty}\dfrac{1}{T}\,\int_0^{T}\psi(\bm{r},t)\,\psi^*(\bm{r},t+\tau)\,dt .
\end{equation}
Selecting different positions $\bm{r}$ of the condensate, we can measure the correlation function for the individual polaritonic eigenmodes 
$\psi_{0,1}(\bm{r})$ and their superposition, as shown in Fig.~\ref{fig:figure5} and detailed in Supp. Mat.~\ref{sec:app:coherence}.
We then fit the experimentally measured coherences for the two eigenmodes with  $|g^{(1)}_{0,1}(\tau)| \simeq e^{- (\tau/\tau_{c0,c1})^2}$ 
and for their superposition with $|g^{(1)}(\tau)| \simeq \frac{1}{2} \big[|g^{(1)}_{0}(\tau)| \cos(2\Omega \tau) + |g^{(1)}_{1}(\tau)|  \big]$
extracting thereby the corresponding coherence times and beating frequency $2\Omega$.  

Consistent with the previous studies \cite{Orfanakis_Ultralong_2021}, we observe that the polariton condensates spatially separated from the trapping potential can exhibit ultralong coherence times. Yet we obtain that the polaritonic eigenmodes $\psi_{0}(\bm{r})$ and $\psi_{1}(\bm{r})$ have different coherence times $\tau_{c0} > \tau_{c1}$. We explain this difference by the proximity of the wavefunction of the corresponding eigenmode to the laser-induced trapping potential which is subject to the exciton number fluctuations and laser noise. The lower energy eigenmode having the two-lobe wavefunction $\psi_{0}(\bm{r})$ well separated from the walls of the confining potential has longer coherence time than the higher energy eigenmode $\psi_{1}(\bm{r})$ whose two lobes have larger overlap with the confining potential and therefore susceptible to more dephasing.


\paragraph{Discussion.}

To summarize, we have experimentally realized and characterized a novel semiconductor exciton-polariton condensate system that exhibits unique properties, including long coherence times and persistent oscillations between two energy-degenerate vortex states $\ket{\circlearrowleft},\ket{\circlearrowright}$ with tunable frequency $\Omega$.
We may argue that our polaritonic system, being equivalent to a spin-1/2 in a magnetic field $\Omega$, is a promising candidate for quantum information applications. The two eigenstates of the polaritonic condensate split by $\mp \Omega$ can serve as the qubit basis states $\ket{0,1} = (\ket{\circlearrowleft} \pm \ket{\circlearrowright})/\sqrt{2}$ equivalent to the spin states $\ket{\uparrow,\downarrow}_z$. The coherent oscillations between the vortex states are equivalent to the spin precession in the $xy$ plane perpendicular to the magnetic field direction along $z$. Using an auxiliary control laser beam, we can control the oscillation frequency and even initialize the qubit to the well-defined state, e.g. $\ket{0}$. Moreover, by changing the position of the sufficiently strong control beam, we can overwrite the asymmetry of the confining potential or the position of the intrinsic defect, which would amount to the rotation of the direction of the effective magnetic field in the $yz$ plane. This will induce coherent spin precession (qubit rotations) in any plane containing the $x$ axis of the Bloch sphere in Fig.~\ref{fig:figure1}. Finally, by stroboscopic application of suitable control beams, we may implement spin-echo or bang-bang operations to freeze the state of the qubit when required. 
Complemented with the controllable interactions between pairs of polaritonic qubits (see Supp. Mat.~\ref{sec:app:tqgates}), such systems can simulate a large subset of quantum computing algorithms that do not rely on entanglement or projective measurements of the genuine qubits \cite{Byrnes2012,Biham2004,Lanyon2008,Balynsky2021,Mohseni2022}.


\paragraph{Acknowledgments.}
J.B., M.N., X.Zh., G.P., and P.G.S. acknowledge the support of Westlake University, Project No. 041020100118 and Program No. 2018R01002 funded by Leading Innovative and Entrepreneur Team Introduction Program of Zhejiang Province of China. A.F.T and D.P. were supported by the EU QuantERA Project PACE-IN (GSRT Grant No. T11EPA4-00015).

\bigskip



\clearpage
\onecolumn
\begin{center}
\textbf{\LARGE Supplementary Material}
\end{center}
\setcounter{equation}{0}
\setcounter{figure}{0}
\setcounter{table}{0}
\setcounter{page}{1}
\makeatletter
\renewcommand{\theequation}{S\arabic{equation}}
\renewcommand{\thefigure}{S\arabic{figure}}

\section{Experiment setup}
\label{sec:app:experiment}

\begin{figure}[h]
    \centerline{\includegraphics[clip,width=0.75\linewidth]{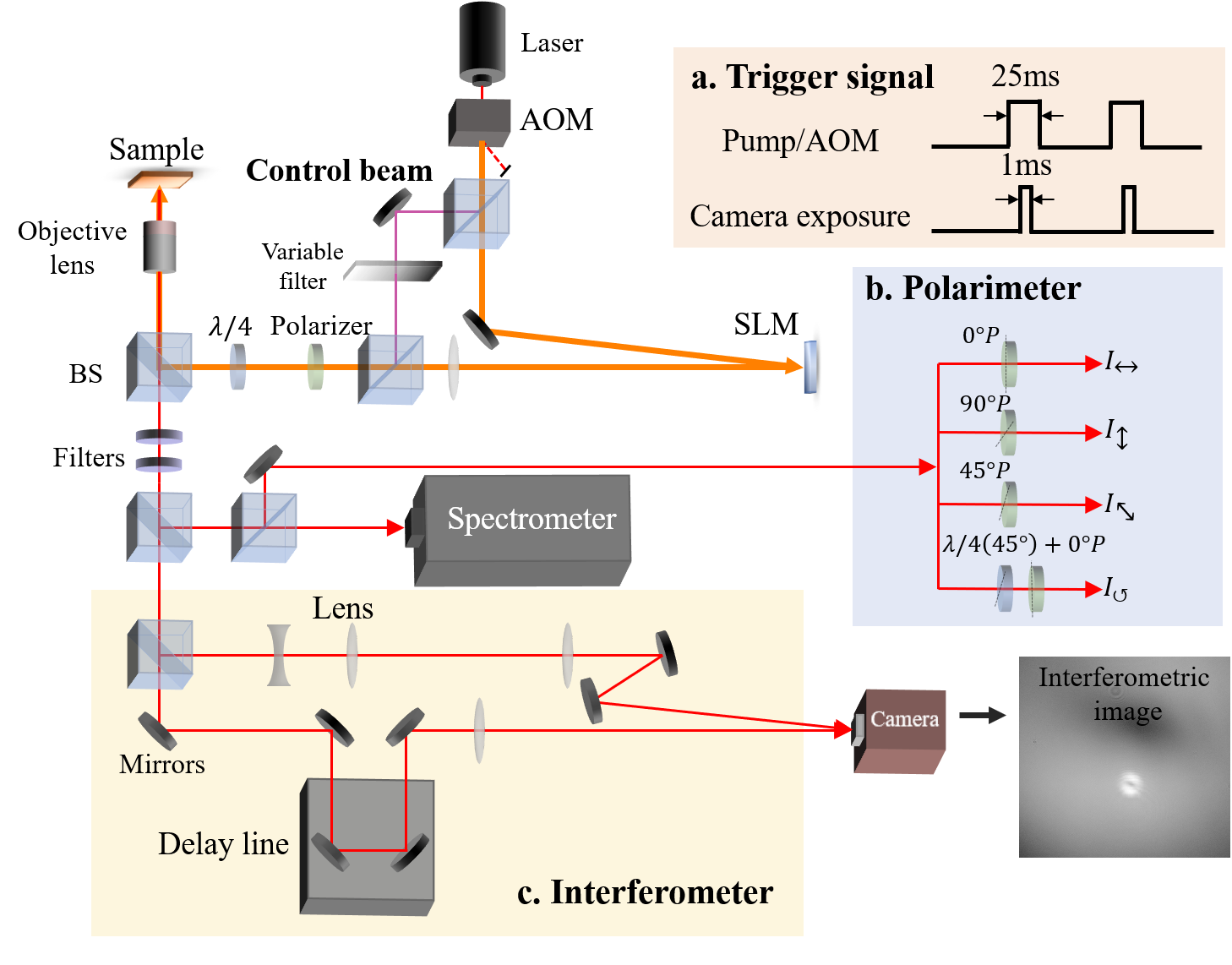}}
    \caption{\textbf{Schematics of the experimental set-up.}
   Main panel: A CW laser is split into two beams. The first (stronger) beam is used for polariton excitation and trapping. SLM generates a Mexican hat-shaped potential for the pump laser. The second beam is used as a weak control beam whose intensity is tuned by a variable neutral density (ND) filter.
   A polarizer and a $\lambda/4$ wave-plate are used to circularly polarize the pump beam, which is then projected onto the sample surface via a microscope objective. The filter blocks the scattered pump laser and the residual background substrate emission. The PL beam is then split into three detection setups for complete characterization of the condensate.  
    \textbf{a.}~The camera, spectrometer and AOM are triggered synchronously, such that the acquired image is averaged over duration of only one pulse. 
   \textbf{b.}~Schematic diagram of the polarimeter. 
    \textbf{c.}~The interferometer employed for the polariton phase and amplitude retrieval comprises of delay line and a magnification optics.
}
    \label{fig:figure6}
\end{figure}

A $5\lambda/2$ GaAs semiconductor microcavity sample\cite{Tsotsis2012S} is kept at a cryogenic temperature of 11K and is excited by a circularly polarized, stabilized continuous wave (CW) laser tuned at 1.6545eV above microcavity  reflection stopband. The laser beam is modulated by an acousto-optic modulator (AOM) to produce $25\,$ms long quasi-CW pulses (much longer than the system dynamics) at a frequency of 4Hz  to reduce the heating of the sample. A measurement window of $1\,$ms (driven by a synchronized trigger) is used to collect the signal from a single pulse realization of a condensate, as shown in Fig.~\ref{fig:figure6}a. Fresnel lens positioned in the incident laser beam transforms the computer-generated hologram imprinted onto a spatial light modulator (SLM) to generate a Mexican-hat excitation profile shown in Fig.~\ref{fig:figure6}. Thus the SLM hologram consists of a combination of Fresnel and Axicon lenses to generate a ring profile. Furthermore, Zernike masks are implemented to compensate for optical aberrations and to control the ring ellipticity. The combination of these masks provides the high degree of control required to generate the nearly defect-free potential. Control beam with adjustable intensity originating from the same laser is overlapped with the original pump beam and projected onto the sample via a microscope objective lens as shown in Fig.~\ref{fig:figure6}.

The polariton condensate emission is collected via the microscope objective. It is filtered to eliminate the laser light and emission from the substrate. This filter is removed to record the composite images of pump excitation profile and PL emission of the condensate shown in Figs.~\ref{fig:figure2}c and \ref{fig:figure4}a. To characterize the condensate emission, we split the PL into three paths leading to a spectrometer, a polarimeter, and an interferometer as shown in Fig.~\ref{fig:figure6}. The specially designed polarimeter analyzes and records on the same camera the images of four polarization components (vertical, horizontal, diagonal, and left-circular) and computes spatially resolved Stokes parameters of the PL as seen in Fig.~\ref{fig:figure6}b. In the interferometer, the beam is split into reference/signal arms by a 90/10 ratio, with the weak signal beam propagating through a delay line and the reference beam being expanded by a factor $\sim 10$ as seen in Fig.~\ref{fig:figure6}c. The two arms are recombined at an angle and imaged onto a camera. Large magnification of the reference beam allows its use as a plane-wave phase reference for phase-retrieval purposes. The resulting interference fringes are processed as described in \cite{Real_Chiral_2021S, Sigurdsson_Persistent_2022S} and the phase and amplitude of the signal are extracted. 
The extracted field $\propto \psi$ is then fit with two vortex modes $\psi_{\circlearrowleft,\circlearrowright}$ having orbital angular momenta $\pm 1$, to compute the normalized vorticities $|\braket{\psi | \circlearrowleft}|^2$ and $|\braket{\psi | \circlearrowright}|^2 \simeq 1-|\braket{\psi | \circlearrowleft}|^2$. By varying the delay between the signal and reference, we measure and fit the periodic coherent oscillation of the polariton condensate between the states $\ket{\circlearrowleft}$ and $\ket{\circlearrowright}$ as shown in Fig.~\ref{fig:figure3}c and map them onto the Bloch sphere in Fig.~\ref{fig:figure3}d.


\section{Interferogram images and their analysis}
\label{sec:app:interference}

Here we describe in more detail the procedure to obtain the interferogram images of Fig.~\ref{fig:figure3} and extract the polariton wavefunction.

To experimentally resolve the dynamics of the system, we perform interferometric measurements of the polariton photoluminescence. The light emanating from the polaritonic condensate $\psi(\bm{r},t)$ is split by an unbalanced beamsplitter and sent into two arms of an interferometer. One arm of the interferometer carries the signal with the spatiotemporal amplitude $A_\mathrm{S}(\bm{r},t) \propto \psi(\bm{r},t) e^{i\bm{k}_\mathrm{S} \cdot \bm{r}}$ where $\bm{k}_\mathrm{S}$ is the wavevector of the signal beam. In the other arm of the interferometer, the beam is expanded, and the field in the vicinity of a specific location of the ring $\bm{r}_0 = (\rho_0,\theta_0)$ is taken as a reference with amplitude $A_\mathrm{R}(\bm{r},t) \propto \psi(\bm{r}_0,t) e^{i\bm{k}_\mathrm{R} \cdot \bm{r}}$ and nearly uniform phase front in the plane perpendicular to the wavevector $\bm{k}_\mathrm{R}$ along the beam propagation direction. 
Before combining with the reference beam on a CCD camera, we sent the signal beam through a variable delay line to introduce time delay $\tau$. The resulting interferometric image upon a long exposure time $T$ is
\begin{equation}
    \langle\, I(\bm{r}, \tau)\, \rangle = \dfrac{1}{T} \int_0^{T} \big|A_\mathrm{S}(\bm{r},t+\tau)+A_\mathrm{R}(\bm{r},t)\big|^2\, dt.
\end{equation}
From the interference images, employing the Fourier analysis \cite{Real_Chiral_2021S, Sigurdsson_Persistent_2022S}, we can extract the magnitude and phase of the wavefunction. 
Performing the two-dimensional Fourier transform of $\langle\, I(\bm{r},\tau)\, \rangle$, we obtain $\mathcal{I}_{\bm{k}}$ with three peaks around $\bm{k}=0,\pm \Delta \bm{k}$, where $\Delta \bm{k}\equiv \bm{k}_\mathrm{R}-\bm{k}_\mathrm{S}$ is the wavevector tilt that determines the spatial separation of the interference fringes.
The zeroth-order peak $\mathcal{I}_0$ at $\bm{k}=0$ quantifies the 
background of the interferometric picture, and
the pair of peaks of $\mathcal{I}_{\pm\Delta\bm{k}}$ at $\bm{k}=\pm \Delta \bm{k}$ originate from the interference of the two amplitudes $A_{\mathrm{S,R}}$. We select an area in the $k$-space around the peak at $-\Delta \bm{k}$, crop the image, and perform the inverse Fourier transform obtaining the condensate wavefunction $\psi(\bm{r},\tau)$ up to some constant phase offset.   

For the theoretical analysis, we begin with our ansatz for the wavefunction of the system undergoing coherent oscillations with frequency $\Omega$ between the two vortex modes:
\begin{equation}
\psi(\bm{r},t)=R(\rho)\Big[\cos\big(\Omega\,t+\phi_0\big) \, e^{+i \theta} +i\sin\big(\Omega\,t+\phi_0\big) \, e^{-i \theta} \Big] , \label{eq:wavefunct}
\end{equation}
where $R(\rho)$ is the radial (donut-shaped) part common for both vortex modes, $\theta$ is the azimuthal angle, and $\phi_0$ is some initial phase of the oscillations. Then, the interferometric image upon long-time integration is
\begin{eqnarray}
    \langle\, I(\bm{r}, \tau)\, \rangle = \lim_{T\to \infty} \dfrac{1}{T} \int_0^{T} \big|A_\mathrm{S}(\bm{r},t+\tau)+A_\mathrm{R}(\bm{r},t)\big|^2\, dt 
    \propto |R(\rho)|^2 + |R(\rho_0)|^2  +2 R(\rho) \,  R(\rho_0) \,  I_{\text{int}} (\tau), \qquad 
\end{eqnarray}
where the interference term 
\begin{eqnarray}
I_{\text{int}} \equiv  \cos\left(\theta-\theta_0\right) \,  \cos\left(\Delta \bm{k} \cdot \bm{r} \right) \, \cos\left(\Omega \,  \tau\right)  
+ \cos\left(\theta+\theta_0\right) \, \sin\left(\Delta \bm{k} \cdot \bm{r} \right) \,\, \sin\left(\Omega \, \tau\right),
\label{eq:Iint}
\end{eqnarray}
oscillates in time $\tau$ with frequency $\Omega$.
In Fig.~\ref{fig:figure3}(a,b) we compare the experimentally and theoretically  obtained interferograms for $\rho_0 \simeq \rho_c$ and $\theta_0 \simeq \pi/4$ revealing excellent agreement.

Next, we expand the $k$-dependent trigonometric functions of Eq.~(\ref{eq:Iint}) into complex exponentials and obtain the Fourier components
\begin{equation}
    \mathcal{I}_{\pm\Delta\bm{k}}= R(\rho)\left[\cos\left(\theta-\theta_0\right) \, \cos\left(\Omega \tau\right) \mp i\,
         \cos\left(\theta+\theta_0\right) \, \sin\left(\Omega \tau\right)\right],
\end{equation}
up to some overall constant factor. 
For $\theta_0=\pi/4$, we then have
\begin{equation}
    \mathcal{I}_{-\Delta\bm{k}}(\bm{r},\tau)=e^{-i\pi/4}\,\psi(\bm{r},t=\tau'),   \quad \tau' = \tau - \frac{\phi_0 -\pi/4}{\Omega}
\end{equation}
and $\mathcal{I}_{-\Delta\bm{k}}=\mathcal{I}_{+\Delta\bm{k}}^*$, with an oscillating amplitude given by
\begin{eqnarray}
\left|\mathcal{I}_{-\Delta\bm{k}}(\bm{r},\tau)\right| = R(\rho)\sqrt{1+\sin\left(2\theta\right)\,\cos\left(2\Omega\tau\right)} 
\end{eqnarray}

The above analysis verifies that by interferometric measurements of the condensate photoluminescence with varying time-delay $\tau$ we indeed reconstruct the dynamics of the system and recover from the experimental measurements the full wavefunction of the polariton condensate exhibiting coherent oscillations between the two vortex modes.   

\begin{figure}[t]    \centerline{\includegraphics[clip,width=0.6\linewidth]{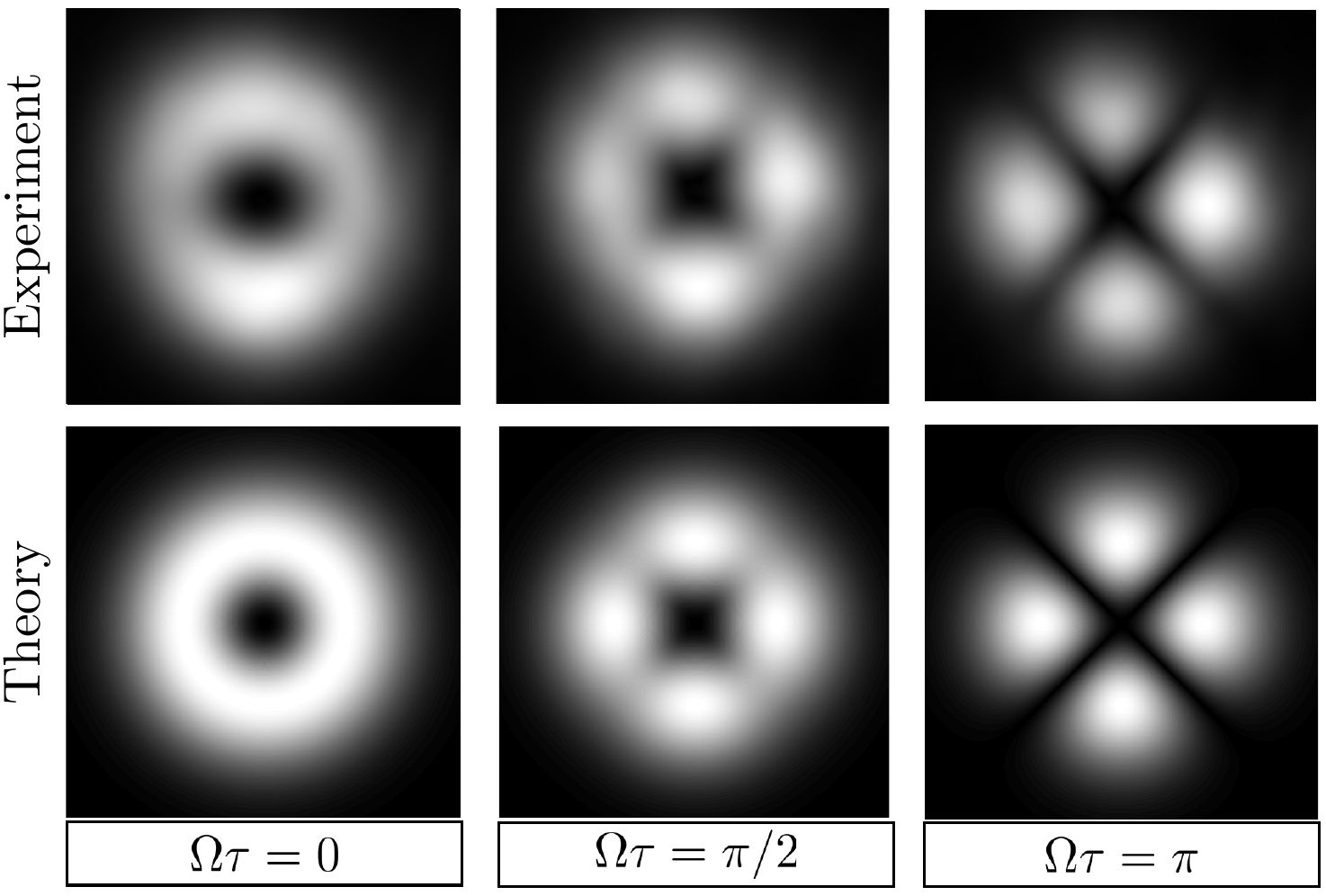}}
    \caption{\textbf{Interference of the condensate PL with itself.} Absolute value of the amplitude of interference of the condensate signal with itself as obtained from the experimental measurements (upper panels) and theoretical model using Eq.~(\ref{eq:Intsws}) (lower panels).}
    \label{fig:Intsws}
\end{figure}

We finally consider the interference of the signal $A_\mathrm{S}(\bm{r},t) \propto \psi(\bm{r},t) e^{i\bm{k}_\mathrm{S} \cdot \bm{r}}$ 
with itself $A_{\mathrm{S}'}(\bm{r},t) \propto \psi(\bm{r},t) e^{i\bm{k}_{\mathrm{S}'} \cdot \bm{r}}$ in a Mach-Zehnder interferometer without expanding the reference
beam. 
Following the same approach as described above, we obtain for the time-averaged intensity 
\begin{eqnarray}
    \langle\, I(\bm{r}, \tau)\, \rangle = \lim_{T\to \infty} \dfrac{1}{T} \int_0^{T} \big|A_\mathrm{S}(\bm{r},t+\tau)+A_{\mathrm{S}'}(\bm{r},t)\big|^2\, dt 
    \propto  2\,|R(\rho)|^2 \big[1 + \cos(\Delta \bm{k} \cdot \bm{r})\cos(\Omega \tau) 
    + \cos(2\theta) \sin(\Delta \bm{k} \cdot \bm{r}) \sin(\Omega \tau) \big] , \label{eq:Intsws}
\end{eqnarray}
where $\Delta \bm{k}\equiv \bm{k}_{\mathrm{S}'}-\bm{k}_\mathrm{S}$.
A comparison between the experimentally retrieved interferometric amplitudes with the theoretical ones obtained using the above formula reveals excellent agreement see Fig.~\ref{fig:Intsws}.

We finally note that the above theoretical derivations assumed fully coherent dynamics of the system, which is valid for delay times $\tau$ much smaller than the coherence time of the condensate $\tau_c$.


\section{Stationary population imbalance of the two eigenmodes}
\label{sec:app:popdiff}

\begin{figure}[t]    
\centerline{\includegraphics[clip,width=0.6\linewidth]{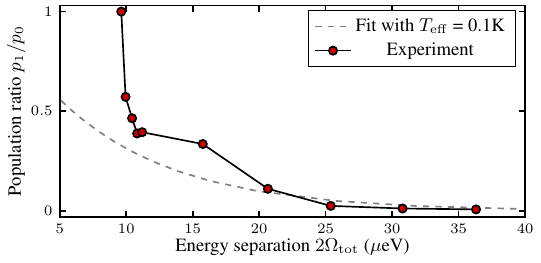}}
    \caption{\textbf{Population imbalance of the two eigenmodes at 11K.} 
    Ratio $p_1/p_0$ of stationary populations $p_{0,1} \equiv |c_{0,1}|^2$ of the two eigenstates $\ket{\uparrow}_\zeta, \ket{\downarrow}_\zeta$ versus their energy separation $2\Omega_{\mathrm{tot}}$ in the presence of a control beam with increasing intensity. The dashed line is the thermal distribution $p_1/p_0= \exp (-2\Omega_{\mathrm{tot}}/k_B T_{\mathrm{eff}})$ with $T_{\mathrm{eff}} \simeq 0.1\;$K }
    \label{fig:Figure9}
\end{figure}

The presence of a strong control beam not only increases the reflection of polariton current and thereby lead to stronger coupling between the two counter-circulating vortex modes but also scatters polaritons in other directions. This scattering, amounting to extra loss, affects the higher energy eigenmode with the antinode at the beam position, while the lower energy eigenmode with a node at the beam position is not affected. As a result of this additional loss for one eigenmode, and not the other, we observe an imbalance in their populations, $p_0 \equiv |c_0|^2 > p_1 \equiv |c_1|^2$, which increases with increasing control beam intensity and thereby their energy separation, as shown in Fig.~\ref{fig:Figure9}. 

We emphasize that the stationary pumping rates of the two eigenmodes are nearly identical and it is the extra loss rate affecting one eigenmode and not the other that causes their population imbalance. Indeed, since the energy separation between the two eigenmodes, $2\Omega_{\mathrm{tot}} \simeq 10-40 \,\mu$eV, is much smaller than the thermal energy $k_{B}T \simeq 0.948\,$meV at our sample temperature $T=11\,$K, this population imbalance cannot be explained by stationary thermal distribution. Attempting to fit the ratio of populations with the Boltzmann distribution, $p_1/p_0 = \exp (-2\Omega_{\mathrm{tot}}/k_B T_{\mathrm{eff}})$, we extract a very low effective temperature $T_{\mathrm{eff}} \simeq 0.1\;$K, while the fit quality is rather poor, which attests to plausibility of our interpretation.


\section{Polariton condensate coherence}
\label{sec:app:coherence}

The Mach Zehnder interferometry of the polariton condensate allows us to determine its coherence time $\tau_c$. 
The visibility of the interference fringes is proportional to the absolute value of the first-order correlation function \cite{Kim_Coherent_2016S}
\begin{equation}
g^{(1)}(\bm{r},\tau) = 
\frac{ \langle \psi(\bm{r},t) \,\psi^*(\bm{r},t+\tau) \rangle}
{\sqrt{\langle |\psi(\bm{r},t) |^2 \rangle 
\langle |\psi(\bm{r},t+\tau) |^2 \rangle}},
\end{equation}
where $\langle \ldots \rangle$ denotes the average over time $t$.
Focusing on the different parts of the condensate, we can then extract the coherence for the two orthogonal eigenmodes $\psi_{0,1}(\bm{r}) = \psi_{\circlearrowleft}(\bm{r}) \pm \psi_{\circlearrowright}(\bm{r})$ and their superposition, see Fig.~\ref{fig:figure5}. 

A polariton condensate spatially separated from the trapping potential can have long coherence times\cite{Orfanakis_Ultralong_2021S}, much longer than the lifetime of individual polaritons. We assume that the main source of coherence relaxation is fluctuations of the trapping potential, which consists of the excitons that repel the polaritons due to the interaction with their excitonic component. Since the excitons are continuously created by the pump laser and annihilated by recombination (or scattering into the polariton mode with smaller probability), their number obeys the Poisson statistics. To extract the condensate coherence, we, therefore, approximate the correlation functions for the two eigenmodes by
\begin{equation}
g^{(1)}_{0,1}(\tau) \approx \exp{\left[\pm i \Omega \tau - (\tau/\tau_{c0,c1})^2\right]} 
\end{equation}
having the Gaussian temporal decay. The correlation function for the superposition of the two modes is then 
\begin{equation}
g^{(1)}(\tau) \approx \frac{1}{2} \left[g^{(1)}_{0}(\tau) + g^{(1)}_{1}(\tau) \right].
\end{equation}

In Fig.~\ref{fig:figure5} we show the experimentally obtained temporal correlation functions for the two condensate modes and their superposition which oscillates with their difference frequency $2\Omega$. Theoretical fits with $|g^{(1)}_{0,1}(\tau)| \simeq e^{- (\tau/\tau_{c0,c1})^2}$ 
and $|g^{(1)}(\tau)| \simeq \frac{1}{2} \left[|g^{(1)}_{0}(\tau)| \cos(2\Omega \tau) + |g^{(1)}_{1}(\tau)|  \right]$ constitute excellent approximation and we extract the oscillation frequency $\Omega\simeq 2\pi\times 4.7\,$ns$^{-1}$ and coherence times $\tau_{c0}\simeq 802.1\,$ps and $\tau_{c1}\simeq 261.7\,$ps.
The vastly different coherence times of the two eigenmodes of the polariton condensate can be explained as follows: 
The lower energy mode $\psi_{0}(\bm{r})$ with the two lobes along the major axis of the elliptical trapping potential is well separated from the fluctuating exciton reservoir and therefore has a long coherence time $\tau_{c0}$. 
On the other hand, the higher energy mode $\psi_{1}(\bm{r})$ has its two lobes along the minor axis of the ellipse and is closer to the shaking trapping potential making it more susceptible to the exciton number fluctuations and the pumping laser noise which leads to increased dephasing and shorter coherence time $\tau_{c1}$.


\section{Analog polaritonic qubits and quantum gates}
\label{sec:app:tqgates}

Our system consists of two degenerate vortex modes $\psi_{\circlearrowleft,\circlearrowright}(\bm{r})$ populated by the polaritonic condensate. Assuming the total number of polaritons $N$, we can represent the state of the two-mode condensate as 
\begin{equation}
\ket{\psi} = \frac{1}{\sqrt{N!}} \big( c_\circlearrowleft a^\dagger_\circlearrowleft + c_\circlearrowright a^\dagger_\circlearrowright\big)^N \ket{\mathbb{0}},
\end{equation} 
where $a^\dagger_{\circlearrowleft,\circlearrowright}$ are the bosonic creation operators for the corresponding modes acting on the condensate vacuum $\ket{\mathbb{0}}$, while the complex amplitudes satisfy $|c_\circlearrowleft |^2 + |c_\circlearrowright |^2 = 1$. Any such state can be mapped onto the Bloch sphere.
We then formally identify the states 
$\ket{\circlearrowleft} = \frac{1}{\sqrt{N!}} (a^\dagger_\circlearrowleft)^N \ket{\mathbb{0}} \to \ket{\uparrow}_x$ 
($c_\circlearrowleft = 1$, $c_\circlearrowright = 0$),
$\ket{\circlearrowright} = \frac{1}{\sqrt{N!}} (a^\dagger_\circlearrowright)^N \ket{\mathbb{0}} \to \ket{\uparrow}_x$ 
($c_\circlearrowleft = 0$, $c_\circlearrowright = 1$); 
$\frac{1}{\sqrt{N!}} \big( \frac{e^{\pm i\pi/4}}{\sqrt{2}} a^\dagger_\circlearrowleft \pm \frac{e^{\mp i\pi/4}}{\sqrt{2}} a^\dagger_\circlearrowright \big)^N \ket{\mathbb{0}} \to \ket{\uparrow,\downarrow}_y$ ($c_{\circlearrowleft,\circlearrowright} = \frac{e^{\pm \pi/4}}{\sqrt{2}}$ and $c_{\circlearrowleft,\circlearrowright} = \frac{e^{\mp \pi/4}}{\sqrt{2}}$); and 
$\frac{1}{\sqrt{N!}} \big( \frac{1}{\sqrt{2}} a^\dagger_\circlearrowleft \pm \frac{1}{\sqrt{2}} a^\dagger_\circlearrowright \big)^N \ket{\mathbb{0}} \to\ket{\uparrow,\downarrow}_z$ ($c_{\circlearrowleft} = \frac{1}{\sqrt{2}}, c_{\circlearrowright} = \pm \frac{1}{\sqrt{2}}$).

In terms of operators $a_{u,d} = (a_\circlearrowleft \pm a_\circlearrowright)/\sqrt{2}$, these states have a more conventional representation:
\begin{gather*}
\ket{\uparrow,\downarrow}_x \equiv \frac{1}{\sqrt{N!}} \left( \frac{1}{\sqrt{2}} a^\dagger_u \pm \frac{1}{\sqrt{2}} a^\dagger_d \right)^N \ket{\mathbb{0}} , \\
\ket{\uparrow}_y \equiv \frac{1}{\sqrt{N!}} \left( \frac{1}{\sqrt{2}} a^\dagger_u + i \frac{1}{\sqrt{2}} a^\dagger_d \right)^N \ket{\mathbb{0}} , \quad
\ket{\downarrow}_y \equiv \frac{1}{\sqrt{N!}} \left( \frac{1}{\sqrt{2}} a^\dagger_d -i \frac{1}{\sqrt{2}} a^\dagger_u \right)^N \ket{\mathbb{0}} , \\ 
\ket{\uparrow,\downarrow}_z \equiv \frac{1}{\sqrt{N!}} \left(a^\dagger_{u,d} \right)^N  \ket{\mathbb{0}} .
\end{gather*}
The Pauli spin operators then correspond to the Schwinder (Stokes) operators 
\begin{align*}
& S_x = a^\dagger_u a_d + a^\dagger_d a_u \to \sigma_x, \\  
& S_y = i a^\dagger_d a_u - i a^\dagger_u a_d \to \sigma_y, \\ 
& S_z = a^\dagger_u a_u - a^\dagger_d a_d \to \sigma_z . 
\end{align*}
The polariton exchange interaction that describes the elastic scattering of the polaritons between the two counter-circulating modes is equivalent, in the spin picture, to an effective magnetic field $\Omega$ in the $z$ direction. The two eigenstates of $S_z \to \sigma_z$, $\ket{\uparrow,\downarrow}_z \to \ket{0,1}$ then form the basis states of our analog qubit.  

Next, we outline two proposals to induce controllable interactions between pairs of polaritonic qubits and realize quantum gates between them. We assume that by properly positioning the control beam, we can induce coherent single-qubit rotations in any plane containing the $z$ axis, as noted in the main text. 

\begin{figure}[t]
\centerline{\includegraphics[clip,width=0.8\linewidth]{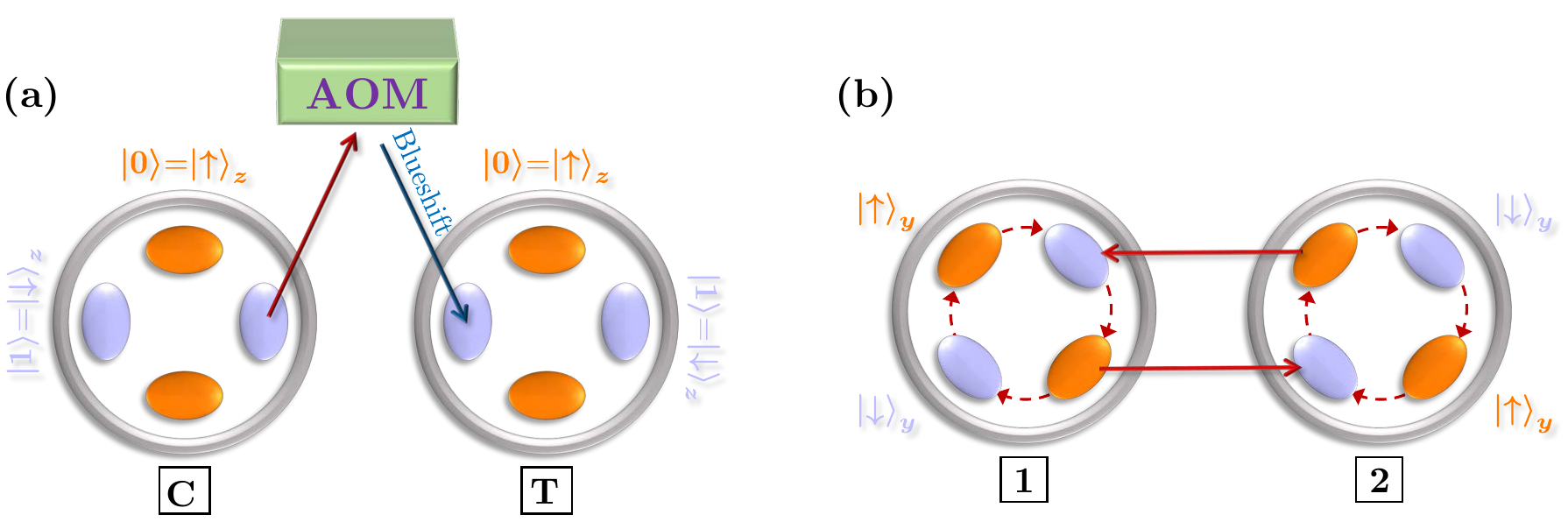}}
    \caption{\textbf{Schemes for realizing two-qubit gates between the polaritonic qubits.} 
    \textbf{a.} The \textsc{cphase} gate between the control (C) and target (T) qubits involves channeling a small fraction of PL from state $\ket{1}$ of the control qubit to AOM and then directing the blue-shifted light onto state $\ket{1}$ of the target qubit to induce conditional potential and thereby energy shift. 
    \textbf{b.} To implement the \textsc{swap} gate between qubits 1 and 2, after rotating the basis states of both qubits, $\ket{0,1} \equiv \ket{\uparrow,\downarrow}_z \to \ket{\uparrow,\downarrow}_y$, 
    we feed a small fraction of PL from state $\ket{\uparrow}_y$ of qubit 1 to state $\ket{\downarrow}_y$ of qubit 2 and vice versa, which, together with the Rabi oscillations $\ket{\uparrow}_y \leftrightarrow \ket{\downarrow}_y$ of each qubit, results in four-wave mixing and effective exchange interaction $\ket{\uparrow}_y \ket{\downarrow}_y \leftrightarrow \ket{\downarrow}_y\ket{\uparrow}_y$ between qubits 1 and 2. After a time required to achieve either \textsc{swap} or $\sqrt{\textsc{swap}}$ operation, we perform the inverse basis rotation $\ket{\uparrow,\downarrow}_y \to \ket{\uparrow,\downarrow}_z$ for both qubits.}
    \label{fig:2qubitgates}
\end{figure}

In the first scheme, see Fig.~\ref{fig:2qubitgates}a, a small fraction of photoluminescence (PL) from state $\ket{1}_\mathrm{C}$ of the control qubit is sent to an acousto-optic modulator (AOM) to up-shift its frequency. The blue-shifted light is then used to induce a small energy shift $\Delta E$ of state $\ket{1}_\mathrm{T}$ of the target qubit. During the interaction time $t_{\mathrm{int}}$, the two-qubit state $\ket{1}_\mathrm{C} \ket{1}_\mathrm{T}$ will accumulate the phase shift $\varphi = -\Delta E t_{\mathrm{int}}$ ($\hbar =1$), while all the other basis states, $\ket{0}_\mathrm{C} \ket{0}_\mathrm{T}, \ket{0}_\mathrm{C} \ket{1}_\mathrm{T}, \ket{1}_\mathrm{C} \ket{0}_\mathrm{T}$, will remain unaffected. At time $t_\mathrm{int} = \pi/\Delta E$, the conditional phase will be $\varphi =-\pi$, which amounts to realizing the controlled-phase, or \textsc{cphase}, gate. This is a universal two-qubit gate equivalent to the \textsc{cnot} gate which can be obtained from it by applying the Hadamarg gates ($\pi/4$ rotation) to the target qubit before and after the  \textsc{cphase} application \cite{David_BookS,Nielsen_BookS}.

In the second scheme, see Fig.~\ref{fig:2qubitgates}b, we consider two qubits, 1 and 2, and first rotate their basis states as $\ket{0,1} = \ket{\uparrow,\downarrow}_z \to \ket{\uparrow,\downarrow}_y$. These states now undergo Rabi oscillations $\ket{\uparrow}_y \leftrightarrow\ket{\downarrow}_y$ with frequency $\Omega$, passing through the vortex states $\ket{\circlearrowleft,\circlearrowright} = \ket{\uparrow,\downarrow}_x$. Next, we take a small fraction of PL from state $\ket{\uparrow}_y$ of qubit 1 and feed it to state $\ket{\downarrow}_y$ of qubit 2 and vice versa. We thus realize four-wave mixing leading to exchange interaction between the qubits 1 and 2 as $\ket{\uparrow}_y \ket{\downarrow}_y \leftrightarrow \ket{\downarrow}_y\ket{\uparrow}_y$. Interrupting this exchange at half circle, we achieve the \textsc{swap} while stopping the oscillations at quarter of the circle we can implement the two-qubit $\sqrt{\textsc{swap}}$ gate
$\ket{\uparrow}_y \ket{\downarrow}_y \to \ket{\uparrow}_y \ket{\downarrow}_y + i \ket{\downarrow}_y\ket{\uparrow}_y$


\begin{thebibliography}{100}


    \bibitem{Deng_Exciton_2010}
    H. Deng, H. Haug, and Y. Yamamoto, Exciton-polariton Bose-Einstein condensation. Rev.Mod. Phys. \textbf{82}, 1489 (2010).
    
    \bibitem{carusotto2013quantum}
    I. Carusotto and C. Ciuti, Quantum fluids of light. Rev. Mod. Phys. \textbf{85}, 299 (2013).

    \bibitem{Deng_Condensation_2002}
    H. Deng, G. Weihs, C. Santori, J. Bloch, and Y. Yamamoto, Condensation of semiconductor microcavity exciton-polaritons. Science \textbf{298}, 199 (2002).

    \bibitem{Kasprzak_Bose_2006}
    J. Kasprzak, et al., Bose–Einstein condensation of exciton polaritons. Nature \textbf{443}, 409 (2006).

    \bibitem{Imamoglu_Nonequil_1996}
    A. Imamoglu, R. J. Ram, S. Pau, and Y. Yamamoto,  Nonequilibrium condensates and lasers without inversion: Exciton-polariton lasers. Phys. Rev. A \textbf{53}, 4250 (1996).

    \bibitem{Christopoulos_Room_2007}
    S. Christopoulos, et al., Room-temperature polariton lasing in semiconductor microcavities. Phys. Rev. Lett. \textbf{98}, 126405 (2007).

    \bibitem{Sanvitto_Theroad_2016}
    D. Sanvitto, and S. Kéna-Cohen, The road towards polaritonic devices. Nat. Mat. \textbf{15}, 1061 (2016). 

    \bibitem{Brien_Photonic_2009}
    J. O'Brien, A. Furusawa, and J. Vučković, Photonic quantum technologies. Nat. Phot. \textbf{3}, 687 (2009). 

    \bibitem{Amo_Exciton_2016}
    A. Amo, and J. Bloch, Exciton-polaritons in lattices: A non-linear photonic simulator, C. R. Physique \textbf{17}, 934 (2016).

    \bibitem{Kim_Exciton_2017}
    N. Y. Kim, and Y. Yamamoto, Exciton-polariton quantum simulators. Quantum Simulations with Photons and Polaritons: Merging Quantum Optics with Condensed Matter Physics, 91-121 (2017).

    \bibitem{juggins_2018} R. T. Juggins, J. Keeling \& M. H. Szymańska, Nature Comm.\textbf{9}, 4062 (2018)
    

	
    \bibitem{Kavokin_split_ring_qubit}
    Y. Xue, et al., Split-ring polariton condensates as macroscopic two-level quantum systems. Phys. Rev. Res. \textbf{3}, 013099 (2021).


    \bibitem{Kavokin_polar_q_comput}
    A. Kavokin, T. C. Liew, C. Schneider, P. G. Lagoudakis, S. Klembt, and S. Hoefling, Polariton condensates for classical and quantum computing. Nat. Rev. Phys. \textbf{4}, 435 (2022).

	
    \bibitem{Mooij_Josephson_1999}
    J. E. Mooij, T. P. Orlando, L Levitov, L. Tian, C. H. Van der Wal, and S. Lloyd, Josephson persistent-current qubit. Science \textbf{285}, 1036 (1999).

    \bibitem{Chiorescu_Coherent_2003}
    I. Chiorescu, Y. Nakamura, C. M. Harmans, and J. E. Mooij, Coherent quantum dynamics of a superconducting flux qubit. Science \textbf{299}, 1869 (2003).
    
    \bibitem{You_Atomic_2011}
    J. Q. You, and F. Nori, Atomic physics and quantum optics using superconducting circuits. Nature \textbf{474}, 589 (2011). 

    \bibitem{Arute_Quantum_2019}
    F. Arute, et al., Quantum supremacy using a programmable superconducting processor. Nature \textbf{574}, 505 (2019).  
    

	
    \bibitem{Lagoudakis_1}
    K. G. Lagoudakis, et al.,  Quantized vortices in an exciton–polariton condensate. Nat. Phys. \textbf{4} 706 (2008).

    \bibitem{Lagoudakis_2}
    K. G. Lagoudakis, et al., Observation of half-quantum vortices in an exciton-polariton condensate. Science \textbf{326}, 974 (2009).

    \bibitem{Manni_Dissociation_2012}
    F. Manni, et al., Dissociation dynamics of singly charged vortices into half-quantum vortex pairs. Nat. Commun. \textbf{3}, 1309 (2012).

    \bibitem{Grosso_Soliton_2011}
    G. Grosso, G. Nardin, F. Morier-Genoud, Y. Léger, and B. Deveaud-Plédran, Soliton instabilities and vortex street formation in a polariton quantum fluid. Phys. Rev. Lett. \textbf{107}, 245301 (2011).

    \bibitem{Nardin_Hydrodynamic_2011}
    G. Nardin, et al., Hydrodynamic nucleation of quantized vortex pairs in a polariton quantum fluid. Nat. Phys. \textbf{7}, 635 (2011).
    
    
    \bibitem{Dall_Creation_2014}
    R. Dall, et al., Creation of orbital angular momentum states with chiral polaritonic lenses. Phys. Rev. Lett. \textbf{113}, 200404 (2014).

    \bibitem{Gao_Chiral_2018}
    T. Gao, et al., Chiral modes at exceptional points in exciton-polariton quantum fluids. Phys. Rev. Lett. \textbf{120}, 065301 (2018).
    
    \bibitem{Gao_Controlled_2018}
    T. Gao, et al., Controlled ordering of topological charges in an exciton-polariton chain. Phys. Rev. Lett. \textbf{121}, 225302 (2018).

    \bibitem{Sedov_Circular_2021}
    E. S. Sedov, V. A. Lukoshkin, V. K. Kalevich, P. G. Savvidis, and A. V. Kavokin, Circular polariton currents with integer and fractional orbital angular momenta. Phys. Rev. Res. \textbf{3}, 013072, (2021).

    \bibitem{Aladinskaia_Spatial_2023}
    E. Aladinskaia, et al., Spatial quantization of exciton-polariton condensates in optically induced traps. Phys. Rev. B \textbf{107}, 045302 (2023).

    
    \bibitem{Lukoshin_Persistent_2018}
    V. A. Lukoshkin, et al., Persistent circular currents of exciton-polaritons in cylindrical pillar microcavities. Phys. Rev. B \textbf{97}, 195149 (2018).
    
    \bibitem{Real_Chiral_2021}
    B. Real, et al., Chiral emission induced by optical Zeeman effect in polariton micropillars. Phys. Rev. Res. \textbf{3}, 043161 (2021).
    

    \bibitem{Sanvitto_Persistent_2010}
    D. Sanvitto, et al., Persistent currents and quantized vortices in a polariton superfluid. Nat. Phys. \textbf{6}, 527 (2010).

    \bibitem{Kwon_Direct_2019}
    M. S. Kwon, et al., Direct transfer of light’s orbital angular momentum onto a nonresonantly excited polariton superfluid. Phys. Rev. Lett. \textbf{122}, 045302 (2019).
    
    \bibitem{Dominici_Shaping_2021}
    L. Dominici, et al. Shaping the topology of light with a moving rabi-oscillating vortex. Opt. Express \textbf{29}, 37262 (2021).
    
    \bibitem{Choi_Observation_2022}
    D. Choi, et al., Observation of a single quantized vortex vanishment in exciton-polariton superfluids. Phys. Rev. B \textbf{105}, L060502 (2022).
    
    
    \bibitem{Ma_Realization_2020}
    X. Ma, et al., Realization of all-optical vortex switching in exciton-polariton condensates. Nat. Comm. \textbf{11}, 1 (2020).
    
    \bibitem{Alyatkin_All}
    S. Alyatkin, et al., All-optical artificial vortex matter in  quantum fluids of light. arXiv:2207.01850 (2022).
    
    
    \bibitem{Manni_Spontaneous_2011}
    F. Manni, K. G. Lagoudakis, T. C. H. Liew, R. André, and B. Deveaud-Plédran, Spontaneous pattern formation in a polariton condensate. Phys. Rev. Lett. \textbf{107}, 106401 (2011).
    
    \bibitem{Dreismann_Coupled_2014}
    A. Dreismann, et al., Coupled counterrotating polariton condensates in optically defined annular potentials. Proc. Natl. Acad. Sci. U.S.A. \textbf{111}, 8770 (2014).
    
    \bibitem{Wang_Spontaneously_2021}
    J. Wang, et al., Spontaneously coherent orbital coupling of counterrotating exciton polaritons in annular perovskite microcavities. Light. Sci. Appl. \textbf{10}, 45 (2021).
    
    \bibitem{Zhang_All_2023}
    S. Zhang, et al., All-optical control of rotational exciton polaritons condensate in perovskite microcavities. ACS Photonics (2023).
    

    \bibitem{Gnusov_Quantum_2023}
    I. Gnusov, et al., Quantum vortex formation in the “rotating bucket” experiment with polariton condensates. Sci. Adv. \textbf{9}, eadd1299 (2023).

    \bibitem{Byrnes2012}
    T. Byrnes, K. Wen, and Y. Yamamoto,
    Macroscopic quantum computation using Bose-Einstein condensates. 
    Phys. Rev. A \textbf{85}, 040306(R) (2012).


    \bibitem{Biham2004}
    E. Biham, G. Brassard, D. Kenigsberg, and T. Mor, 
    Quantum computing without entanglement. 
    Theor. Comput. Sci. \textbf{320}, 15 (2004).

    \bibitem{Lanyon2008}
    B. P. Lanyon, M. Barbieri, M. P. Almeida, and A. G. White, Experimental quantum computing without entanglement. 
    Phys. Rev. Lett. \textbf{101}, 200501 (2008).

    \bibitem{Balynsky2021}
    Balynsky, M. et al.,
    Quantum computing without quantum computers: Database search and data processing using classical wave superposition. 
    J. Appl. Phys. \textbf{130}, 164903 (2021).
    
    \bibitem{Mohseni2022}
    M. Mohseni, et al.,
    Classical analog of qubit logic based on a magnon Bose–Einstein condensate, 
    Communications Phys. \textbf{5}, 196 (2022).
    
	
    \bibitem{Mukherjee_Dynamics_2021}
    S. Mukherjee, et al., Dynamics of spin polarization in
    tilted polariton rings. Phys. Rev. B. \textbf{103},  165306 (2021).
    
    \bibitem{Liu_Anew_2015}
    G. Liu, D. W. Snoke, A. Daley, L. N. Pfeiffer, and K. West, A new type of half-quantum circulation in a macroscopic polariton spinor ring condensate. Proc. Natl. Acad. Sci. U.S.A. \textbf{112}, 2676 (2015).
    
    \bibitem{Dominici_Vortex_2015}
    L. Dominici, et al., Vortex and half-vortex dynamics in a nonlinear spinor quantum fluid. Sci. Adv. \textbf{1}, e1500807 (2015). 
    
    \bibitem{Gnusov_All_2021}
    I. Gnusov, et al., All-optical linear-polarization engineering in single and coupled exciton-polariton condensates. Phys. Rev. Appl. \textbf{16}, 034014 (2021).
    
    \bibitem{Pukrop_Circular_2020}
    M. Pukrop, S. Schumacher, and X. Ma, Circular polarization reversal of half-vortex cores in polariton condensates.  Phys. Rev. B \textbf{101}, 205301 (2020).
    
    \bibitem{Demirchyan_Spinor_2022}
    S. Demirchyan, I. Chestnov, K. Kondratenko, and A. Kavokin, Spinor superfluid currents of exciton-polaritons on a split-ring. arXiv:2208.14094 (2022).
    
    \bibitem{Aristov_Screening_2022}
    D. Aristov, H. Sigurdsson, and P. G. Lagoudakis, Screening nearest-neighbor interactions in networks of exciton-polariton condensates through spin-orbit coupling. Phys. Rev. B. \textbf{105}, 155306 (2022).
    
    \bibitem{Gnusov_Optical_2020}
    I. Gnusov, et al., Optical orientation, polarization pinning, and depolarization dynamics in optically confined
    polariton condensates. Phys. Rev. B. \textbf{102}, 125419 (2020).  


    \bibitem{Sigurdsson_Persistent_2022}
    H. Sigurdsson, et al., Persistent self-induced Larmor precession evidenced through periodic revivals of coherence.	Phys. Rev. Lett. \textbf{129}, 155301 (2022).
    
    
    \bibitem{Orfanakis_Ultralong_2021}
    K. Orfanakis, A. F. Tzortzakakis, D. Petrosyan, P. G. Savvidis, and H. Ohadi, Ultralong temporal coherence in optically trapped exciton-polariton condensates. Phys. Rev. B. \textbf{103}, 235313 (2021).

\end{thebibliography}

\bigskip

\begin{thebibliography}{11}  
    
   \bibitem[S1]{Tsotsis2012S}
    P. Tsotsis, et al., Lasing threshold doubling at the crossover from strong to weak coupling regime in GaAs microcavity. New J. Phys. \textbf{14}, 023060 (2012).

    
    \bibitem[S2]{Sigurdsson_Persistent_2022S}
    H. Sigurdsson, et al., Persistent self-induced Larmor precession evidenced through periodic revivals of coherence.	Phys. Rev. Lett. \textbf{129}, 155301 (2022).

    \bibitem[S3]{Real_Chiral_2021S}
    B. Real, et al., Chiral emission induced by optical Zeeman effect in polariton micropillars. Phys. Rev. Res. \textbf{3}, 043161 (2021).

    
    \bibitem[S4]{Kim_Coherent_2016S}
    S. Kim, et al., Coherent polariton laser. Phys. Rev. X \textbf{6},  011026 (2016).
     
    \bibitem[S5]{Orfanakis_Ultralong_2021S}
    K. Orfanakis, A. F. Tzortzakakis, D. Petrosyan, P. G. Savvidis, and H. Ohadi, Ultralong temporal coherence in optically trapped exciton-polariton condensates. Phys. Rev. B. \textbf{103}, 235313 (2021).

    
    \bibitem[S6]{David_BookS}
    P. Lambropoulos, and D. Petrosyan, Fundamentals of Quantum Optics and Quantum Information.  Springer, Berlin (2007).
    
    \bibitem[S7]{Nielsen_BookS}
    M. A. Nielsen, and I. L. Chuang, Quantum Computation and Quantum Information. Cambridge University Press, New York (2000).    
\end{thebibliography}
\end{document}